\documentclass[twocolumn,10pt,prl, aps,superscriptaddress,showpacs]{revtex4-1}
\usepackage{subfigure}

\usepackage{dcolumn}
\usepackage{amsmath,amssymb}
\usepackage{bm}
\usepackage{color}
\usepackage{latexsym}
\usepackage{psfrag,graphicx}
\graphicspath{{figures/}}
\usepackage{color}
\usepackage[english]{babel}
\usepackage{latexsym}

\usepackage{epsf} 
\usepackage{amsfonts}
\usepackage{natbib}
\usepackage{epstopdf}
\usepackage{appendix}
\usepackage{changes}

\definecolor{mygrey}{gray}{0.35}
\definecolor{myblue}{rgb}{0.2,0.2,0.8}
\definecolor{myzard}{cmyk}{0,0,0.05,0}
\definecolor{mywhite}{rgb}{1,1,1}
\definecolor{myred}{rgb}{1,0.,0.3}

\usepackage[colorlinks=true,citecolor=myblue,linkcolor=myred]{hyperref}

\def\be{\begin{equation}}
\def\ee{\end{equation}}
\def\ba{\begin{align}}
\def\enda{\end{align}}
\def\bi{\begin{itemize}}
\def\ei{\end{itemize}}

\def\beq{\begin{equation}}
\def\beq{\begin{equation}}
\def\eeq{\end{equation}}
\def \bea{\begin{eqnarray}}
\def \eea{\end{eqnarray}}

 \def\ee{\mathord{\rm e}}

\renewcommand{\ee}{{\rm e}}

\newcommand{\ket}[1]{|#1\rangle}
\newcommand{\bra}[1]{\langle #1|}

\newcommand{\bla}[1]{\left(#1\right)}
\newcommand{\blb}[1]{\left[#1\right]}

\begin{document}

\title[Short Title]{Fast dynamical decoupling of the M{\o}lmer-S{\o}rensen entangling gate}

\author{Tom Manovitz} \address{Department of Physics of Complex Systems, Weizmann Institute of Science, Rehovot 7610001, Israel}
\author{Amit Rotem} \address{Racah Institute of Physics, The Hebrew University of Jerusalem, Jerusalem 91904, Givat Ram, Israel}
\author{Ravid Shaniv} \address{Department of Physics of Complex Systems, Weizmann Institute of Science, Rehovot 7610001, Israel}
\author{Itsik Cohen} \address{Racah Institute of Physics, The Hebrew University of Jerusalem, Jerusalem 91904, Givat Ram, Israel}
\author{Yotam Shapira}
\author{Nitzan Akerman}
\address{Department of Physics of Complex Systems, Weizmann Institute of Science, Rehovot 7610001, Israel}
\author{Alex Retzker}
\address{Racah Institute of Physics, The Hebrew University of Jerusalem, Jerusalem 91904, Givat Ram, Israel}
\author{Roee Ozeri}
\address{Department of Physics of Complex Systems, Weizmann Institute of Science, Rehovot 7610001, Israel}

\date{June 9th, 2017}


\begin{abstract}
{Engineering entanglement between quantum systems often involves coupling through a bosonic mediator, which should be disentangled from the systems at the operation's end. The quality of such an operation is generally limited by environmental and control noise. One of the prime techniques for suppressing noise is by dynamical decoupling, where one actively applies pulses at a rate that is faster than the typical time scale of the noise.  However, for boson-mediated gates, current dynamical decoupling schemes require executing the pulses only when the boson and the quantum systems are disentangled. This restriction implies an increase of the gate time by a factor of $\sqrt{N}$, with $N$ being the number of pulses applied. Here we propose and realize a method that enables dynamical decoupling in a boson mediated system where the pulses can be applied while spin-boson entanglement persists, resulting in an increase in time that is at most a factor of $\frac{\pi}{2}$, independently of the number of pulses applied. We experimentally demonstrate the robustness of our fast dynamically decoupled entangling gate to $\sigma_z$ noise with ions in a Paul trap.} 
\end{abstract}
                                          
\maketitle

High quality on-demand generation of entanglement is a necessary condition for quantum information processing and quantum metrology. While for some physical platforms entanglement is generated by an inherent direct interaction between subsystems, various platforms of interest make use of a mediating boson with spin-dependent coupling. For instance, the interaction between trapped ions is carried via a vibrational phonon \cite{Molmer1999prl,Molmer2000pra,Wineland2000N,Blatt2014NPH,Monroe2009PRL,Ozeri2014PRA,Tan2, Didi2003nature,Lucas2015Nature1,Lucas2016prl,harty2016prl,Bermudez2012PRA,Tan2013PRL,Itsik2015NJP};  superconducting qubits are entangled via a microwave photon \cite{Zhu2005prl,Majer2007nature,Gao}; the interaction between distant NVs can be carried via a nanomechanical oscillator's phonon \cite{Bennett2013prl,Alex2013njp} and a cavity photon carries the interaction between atoms in cavity QED architectures \cite{Duan,Chou,Solano2003prl}.
The common Hamiltonian representing these quantum systems is of the form
\beq H\left(t\right) = \tilde{\Omega} \sum_i \sigma_{\alpha, i} \bla{b^\dagger e^{i \varepsilon t} +h.c.}, \label{MS}\eeq
with $\sigma_{\alpha, i}$ representing the Pauli matrix in the $\alpha$ direction of the $i^{th}$ spin. 
In the trapped ion case this Hamiltonian allows one to execute the M{\o}lmer-S{\o}rensen (MS) gate \cite{Molmer1999prl}.
 After times $2\pi n/\varepsilon$ for an integer $n$, the boson is disentangled from the spins, leaving the spins entangled via a geometric phase which is proportional to the area of the closed circle traced by the boson trajectory in phase space \cite{Molmer2000pra,Milburn,Roos1}.

Despite considerable progress in achieving high-fidelity entanglement in recent years, entanglement fidelity remains a primary obstacle for performance of large scale quantum information processing, and more particularly fault-tolerant quantum computation. Attempts to improve the fidelity of entangling gates must overcome the limitations imposed by environmental noise as well as imperfections in the control apparatus. Dynamical decoupling is a common method for fighting the effects of noise. When utilizing dynamical decoupling pulses \cite{Hanh1950,Viola1998pra} during the entangling gate operation, one is required to consider the affect on the spin dependent coupling to the mediating boson. In many experiments, a single dynamical decoupling pulse has been applied at a time $2\pi /\varepsilon$, exactly when the boson is disentangled from the spins \cite{Lucas2015Nature1,Lucas2016prl,Tan2013PRL,harty2016prl}.
However,  a single pulse only eliminates the effect of the constant (DC) part of the noise and does not efficiently combat finite frequency (AC) noise. 
Thus, in order to increase the decoupling efficiency, the number of pulses should be increased. Such an increase, however, comes at a price: by increasing the number of dynamical decoupling pulses to $N$, the time interval between the pulses should be shortened by a factor of $\sqrt{N}$, and not $N$ as in NMR schemes \cite{DEER1,DEER2,DEER3}. Hence, the gate duration is prolonged by the same $\sqrt{N}$ factor, or by a factor of $\frac{1}{\tau},$ where $\tau$ is the time between two consecutive pulses. This prolonged time makes the gate more vulnerable to other uncompensated noise sources that reduce the gate fidelity.

The difficulty of adding dynamical decoupling pulses during gate operation occurs when there is a need to apply the pulses in an orthogonal direction to the gate operator. This need often originates from the existence of noise that is parallel to the gate operator, such as the external parallel noise terms in the microwave gradient scheme \cite{Alex2013njp,Duan,Mintert} and in the single sideband protocols in the different platforms \cite{Bermudez2012PRA,Tan2013PRL,Itsik2015NJP,Zhu2005prl,Majer2007nature,Gao,Bennett2013prl,Roos1,Solano2003prl,Chou}. Note that in the case of a slow noise term that is orthogonal to the gate operator, it is sufficient to perform a small number of dynamical decoupling pulses along the direction of the gate operator. Since these pulses commute with the gate operation, they can be applied even when the spins and motion are entangled and therefore without affecting the structure of the gate or its duration. However, when the orthogonal noise is fast, and many parallel dynamical decoupling pulses are needed, the parallel pulse noise accumulates to an appreciable effect. In other words, this parallel pulse noise enforces the use of additional orthogonal dynamical decoupling pulses (like an XY4 or an XY8 sequence) that create the difficulty.

In this paper we present, and experimentally demonstrate with trapped ions, a dynamical decoupling scheme for boson-mediated systems that yields a refocused entangling gate, whose gate duration is increased by a reduced factor of $\sim\pi/2$. This scheme enables implementation of complex dynamical decoupling pulse sequences such - as CPMG and XY8 - in boson-mediated systems. Furthermore, this scheme of using pulsed dynamical decoupling while operating in the fast gate regime can be used with pulsed schemes for suppressing phonon coupling at the end of the gate, especially when more boson modes are involved \cite{Hayes2012prl,Green2015prl,Ivanov2015pra}.

The second order Magnus expansion of the MS Hamiltonian (Eq.~\ref{MS}), obtained from either the MS gate (SI) or the single sideband gate (SI), is the exact solution of the MS unitary,
\begin{multline}
U_{MS}\left(t\right) = \mathcal{D}\blb{\frac{\tilde{\Omega}}{\varepsilon}\sum_{i=1,2}\sigma_{\alpha,i}\left(1-e^{i\varepsilon t}\right)} \cdot\\
\exp \blb{ i\bla{\frac{\tilde{\Omega}}{\varepsilon}\sum_{i=1,2}\sigma_{\alpha,i}  }^{2}\bla{\varepsilon t-\sin\bla{\varepsilon t}} },
\label{unitary}
\end{multline}
with $\alpha=y,x$ denoting the MS or the single sideband gate respectively. 
$\mathcal{D}$ is the displacement operator; therefore the first term traces a circle in phase space with a radius which is proportional to 
$\frac{\tilde{\Omega}}{\varepsilon}.$
In times $t_n = 2\pi n /\varepsilon,$ for an integer $n,$ the system returns to its original location in phase space, meaning the qubits and the boson mediator are disentangled and a pure two qubit state is achieved. The entanglement of this state is proportional to the phase accumulated by the phase space rotations. A maximally entangled gate is generated when the accumulated Berry phase is $4{\tilde\Omega}^2 t_n /\varepsilon = \pi/2$, at a gate duration of $T_{tot} = \pi \sqrt{N}/2\tilde\Omega = \frac{\pi}{8}\varepsilon/\tilde\Omega^2$. Hence the time overhead which is needed for compensating for smaller phase space rotation scales as one over the time of each rotation or as the square root  of the number of pulses. 

\begin{figure}
\subfigure[]{\def \svgwidth{.469 \columnwidth} 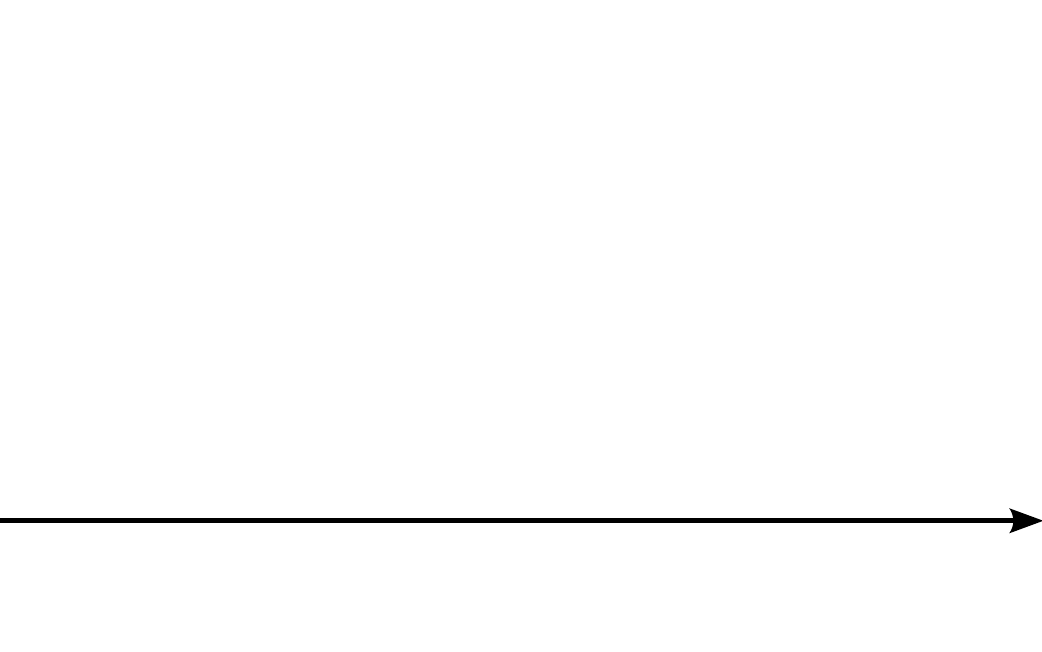 \label{DD at center}}
\subfigure[]{\def \svgwidth{.469 \columnwidth} 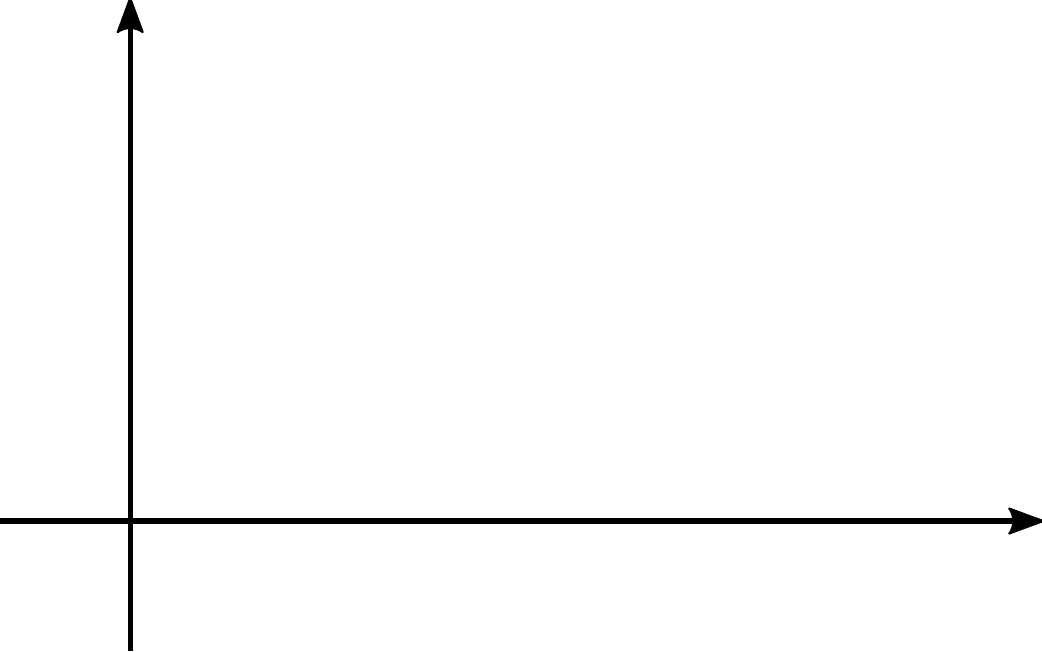 \label{DD as SDK}}
\caption{Trajectory in phase space of a spin with initial state $\ket{\hspace{-4px} \uparrow}$ due to the MS unitary (Eq.~\ref{unitary}).
(A). Applying a $\pi$ pulse when the spin and the boson are disentangled, at times $t_n = 2 \pi n / \varepsilon$, changes between the blue (solid line) and orange (dashed line) trajectories.
(B). By applying a $\pi$ pulse at time $t=\pi/\varepsilon$ the ion trajectory changes from the blue dashed line to the orange dashed line. After a second $\pi$ pulse at time $t=2\pi/\varepsilon$ the ion trajectory changes to the purple line, resulting in a effective $4\tilde{\Omega}/\varepsilon$ displacement of the ions trajectory.}
\end{figure}

The application of dynamical decoupling pulses while implementing a two qubit gate was first developed in the context of NMR architectures \cite{DEER1,DEER2}. In the case where the coupling is $g \sigma_{\alpha,1} \sigma_{\alpha,2}$, applying a $\pi$ pulse dynamical decoupling sequence in a perpendicular direction to $\alpha$ compensates for the single body noisy terms and does not affect the coupling term. This, however, is no longer true when the interaction is mediated by another bosonic degree of freedom, as the pulse sequence might decouple the qubit from the boson mediator or couple it in an uncontrolled way. Hence, the dynamical decoupling pulses are applied when the qubits and the boson mediator are disentangled (Fig.~\ref{DD at center}), i.e.,  at times $t_n = 2 \pi n/ \varepsilon$.
However, to compensate for high frequency noise, a large number of pulses $N \gg 1$ should be applied, resulting in a prolonged overall gate duration by a factor of $\sqrt{N}$  and pulse time separation of $\Delta t \bla{N} = \pi /2 \tilde{\Omega} \sqrt{N}$. Even though the frequency of the DD pulses is increased, thus countering more of the noise spectrum, prolonging the gate duration causes the rest of the noise spectrum to be more damaging to the overall fidelity. Such a scheme is only effective for power spectrums that decay faster than $\frac{1}{\omega}$, meaning the decoherence time will scale as $\frac{S_{BB}(\omega)}{\omega}$ instead of $S_{BB}(\omega)$.

Here we propose an alternative approach in which the dynamical decoupling pulses are used for covering a larger area in the boson phase space. In this way higher boson states are populated during the gate, which is, therefore, performed in the fast gate regime \cite{Molmer2000pra}, regardless of the number of dynamical decoupling pulses involved. The $\pi$ pulses alternate the sign of the $\sigma_{\alpha,i}$ operators in the MS unitary (Eq.~\ref{unitary}), resulting in an effective spin dependent displacement. For instance, a MS unitary for $t = \pi / \varepsilon$ duration gives rise to a spin dependent displacement of $2\tilde{\Omega}/\varepsilon$, and a sequence of MS unitary for $t = \pi / \varepsilon$ duration, $\pi$ pulse, MS unitary for the same $t = \pi / \varepsilon$ duration, and another $\pi$ pulse, results in a double spin dependent displacement of $4\tilde{\Omega}/\varepsilon$ (Fig.~\ref{DD as SDK}).

\begin{figure}
\subfigure[]{\def \svgwidth{.469 \columnwidth} 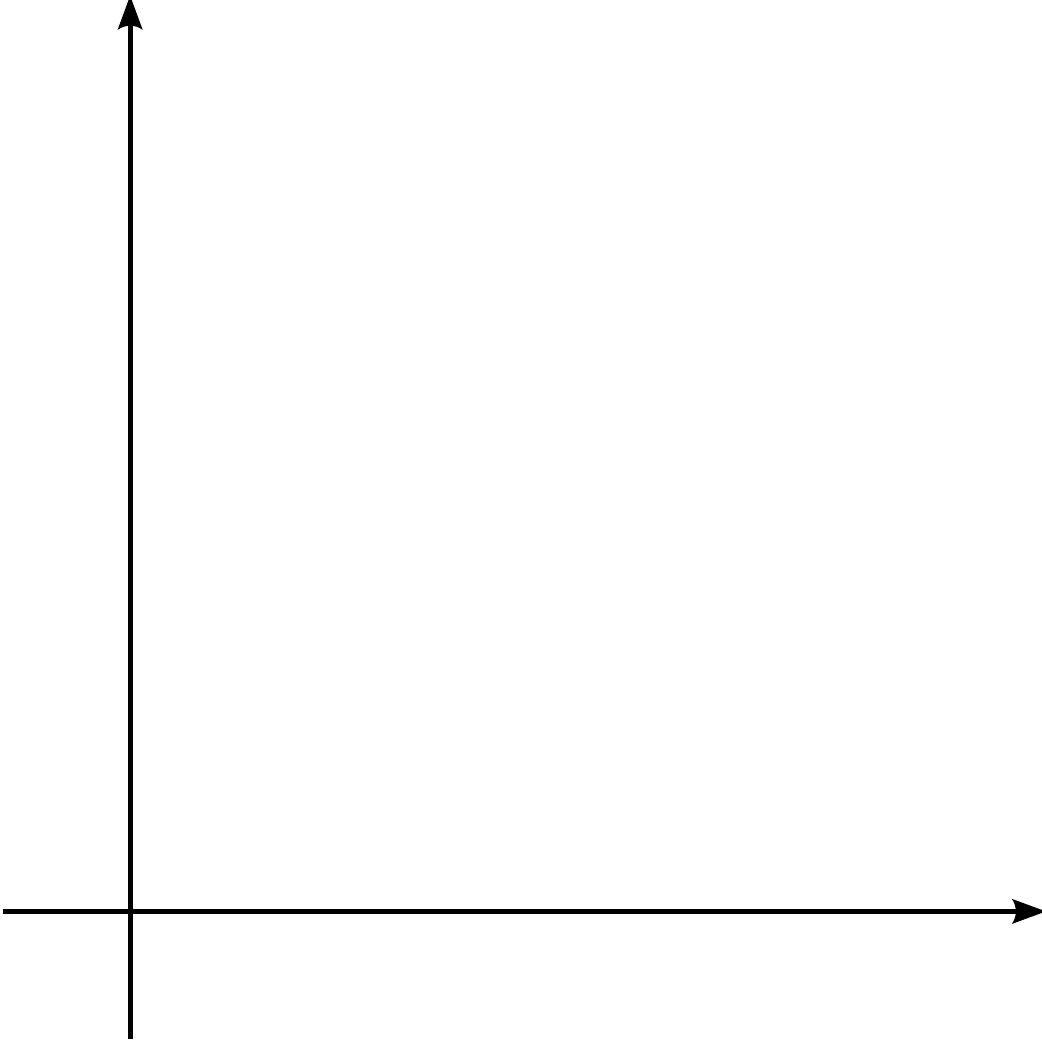 \label{SDK scheme}}
\subfigure[]{\def \svgwidth{.469 \columnwidth} 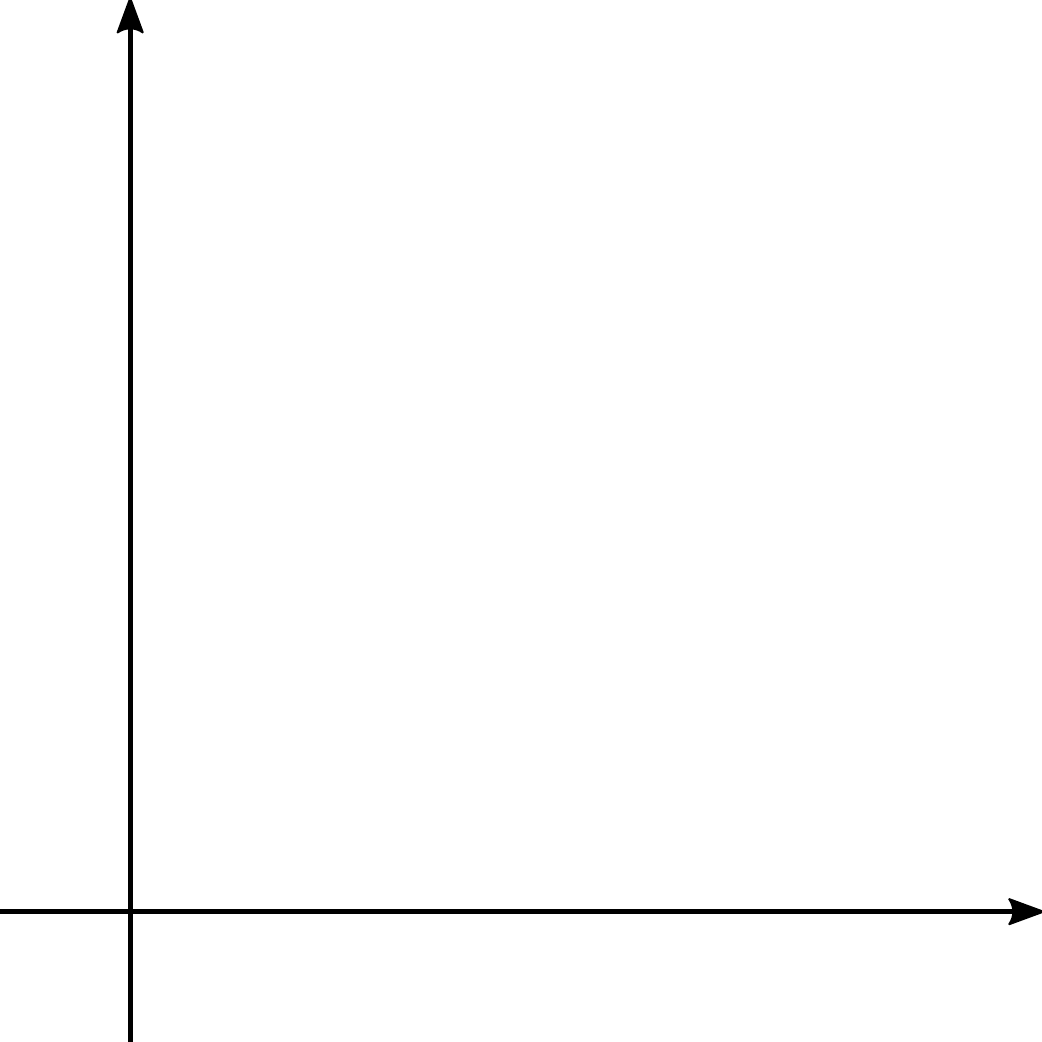 \label{gate speed}}
\caption{(A). Utilizing dynamical decoupling pulses for enlarging the area in phase space. A flower shaped area is generated by applying $N$ dynamical decoupling pulses with a specific time separation $\Delta t(N)={\pi}\left({2} +{N}\right)/N{\varepsilon}$. The area enclosed by the polygon connecting all the turning point, gives rise to the Berry phase  $A = 8 N \left({\tilde{\Omega}}/{\varepsilon}\right)^2 \cot \bla{{\pi}/{N}}$. At the limit of many pulses $1\ll N$, the enlarged area compared to the area associated with the flower's leaves ${4 \tilde{\Omega}^2}T_{tot}/{\varepsilon}$, is responsible for operating in the strong coupling regime.
(B).  In the case where  $1\ll N$, we can neglect the phase accumulated by the flower's leaves. Thus, every two $\pi$ pulses separated by $\Delta t\approx \pi/\varepsilon$ give rise to a spin dependent displacement (orange dashed line) $2\tilde{\Omega}/\varepsilon$, where the path velocity is $2\tilde{\Omega}/\pi$. In comparison to the regular strong coupling entangling gate where the path velocity is $\tilde{\Omega}$ (Eq.~\ref{MS}) we find a factor of $\pi/2$ in the gate durations.
}
\end{figure}

The above method of spin dependent displacements allows for ultrafast gates \cite{Mizrahi2014apb,Garcia2003prl}. By applying $N$ equally-spaced $\pi$ pulses, with time separation $\Delta t(N)={\pi}\left({2} +{N}\right)/N{\varepsilon}$, a flower-shaped path in phase space is closed at the end of the gate operation $T_{tot}  = {\pi \left(N+2\right)}/{\varepsilon}$, thus remaining decoupled from the boson field(Fig.~\ref{SDK scheme}). The Berry phase accumulated in the area enclosed by the flower's petals  ${4 \tilde{\Omega}^2}T_{tot}/{\varepsilon} $ is equal to the Berry phase accumulated in the slow-coupling regime, i.e. the area accumulated without applying the dynamical decoupling pulses. However, by applying the dynamical decoupling pulses an additional Berry phase  is accumulated in the polygon area $A = 8 N \left({\tilde{\Omega}}/{\varepsilon}\right)^2 \cot \bla{{\pi}/{N}}$.

A maximally entangled state can be generated when the overall accumulated Berry phase is 
\beq
A + \frac{4 \tilde{\Omega}^2}{\varepsilon} T_{tot} =  \left(\frac{2\tilde{\Omega}}{\varepsilon}\right)^2 \left(\varepsilon T_{tot} + 2 N \cot \frac{\pi}{N} \right) {=} \frac{\pi}{2},
\eeq
with the gate duration being a monotonically increasing function of $N$
\beq
T_{tot} = \frac{\pi}{2 \tilde{\Omega}} \frac{\frac{N}{2}+1}{\sqrt{\frac{N}{2}+1 + \frac{N}{\pi} \cot\frac{\pi}{N}}} \underset{N \rightarrow \infty}{\rightarrow} \frac{\pi}{2 \tilde{\Omega}} \frac{\pi}{2},
\eeq
 and with a pulse time separation limit $\Delta t \bla{N} \rightarrow \pi^2/4 \tilde{\Omega} N $.
Hence, although applying a large number of dynamical decoupling pulses $N\gg 1$, the gate duration is prolonged only by a factor of less than $\pi/2$. 
 Intuitively, this can be understood by the following observation: at this limit of $N\gg 1$ the Berry phase accumulated in the flower's leaves is negligible relative to the polygon phase, which is approximately a circle. Every separation time of $\Delta t\approx \pi/\varepsilon$, we accumulate a $2\tilde{\Omega}/\varepsilon$ spin dependent displacement  (Fig.~\ref{gate speed}). Therefore, the displacement velocity in phase space, which is the angular velocity of the accumulated circle, is $2\tilde{\Omega}/\pi$. Comparing this to the regular strong coupling gate, where the displacement velocity is $\tilde{\Omega}$ (Eq.~\ref{MS}), and taking into account that both gates should accumulate the same circle area in phase space, we find a factor of $\pi/2$ in the gate durations. In this argument we omitted the area associated with the flower's leaves, which is increased when $N$ decreases. Thus, for a smaller number of pulses, the area accumulated by the polygon should be reduced, and the gate duration is shorter.  

Note that in our derivation we have considered instantaneous dynamical decoupling pulses, which is justified when the spin-boson interaction and trap dynamics are negligible during the pulse. Similar results can be achieved by turning off the driving field that is responsible for the coupling, as in our experimental realization discussed below, or by increasing the detuning during the pulses, in which case only the trap dynamics needs reconsideration. In the former case, the pulse sequence $\left\{\varepsilon t_n \right\}_n$ remains the same, and the detuning ($\varepsilon$) requires adjustment on the order of $\bla{\Omega \eta \tau}^2$, where $\tau$ is the pulse duration. 

\begin{figure*}  
\subfigure[]{\includegraphics[width=0.66 \columnwidth]{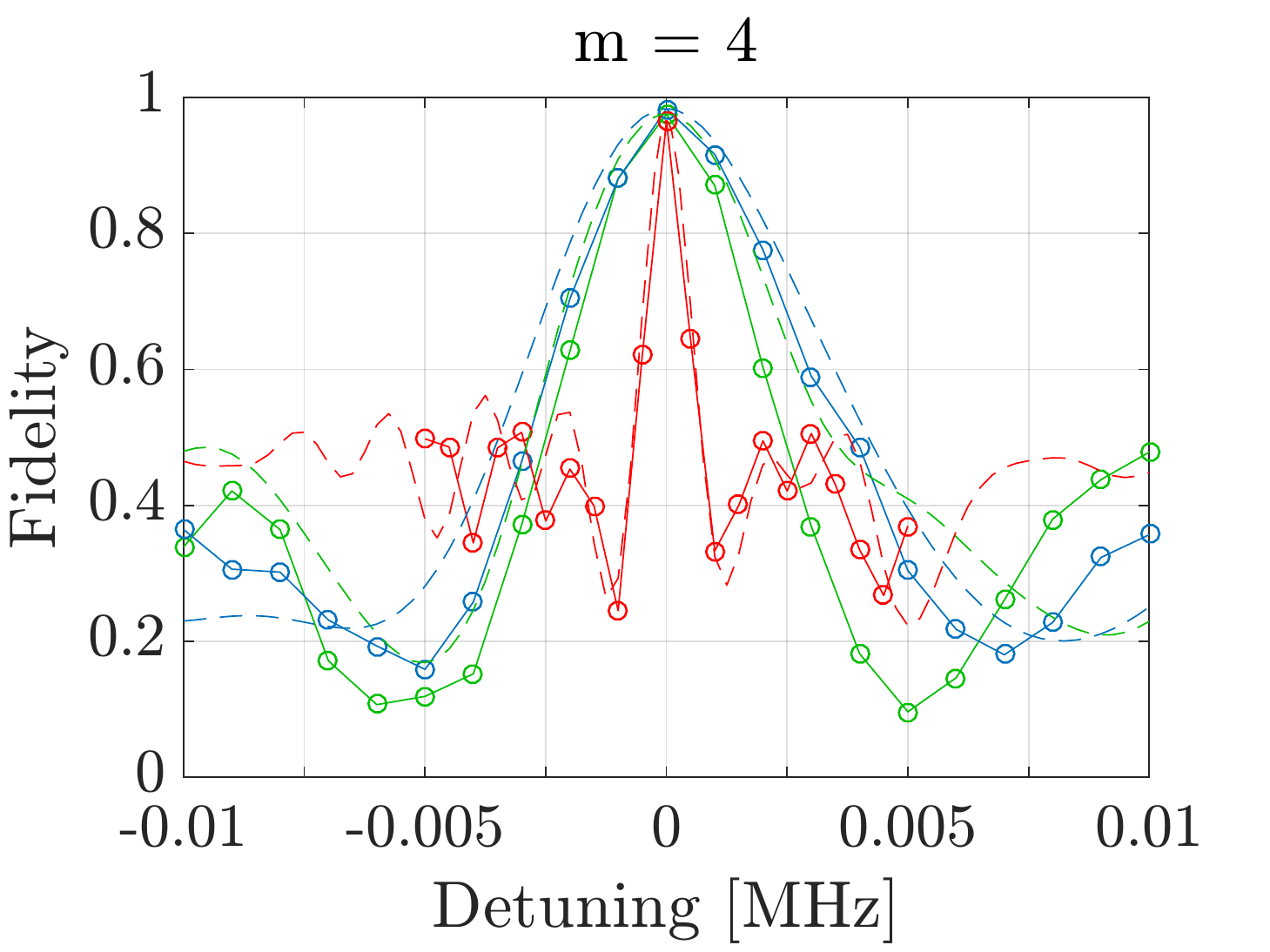} \label{4 pulses Fidelity vs detuning}}
\subfigure[]{\includegraphics[width=0.66 \columnwidth]{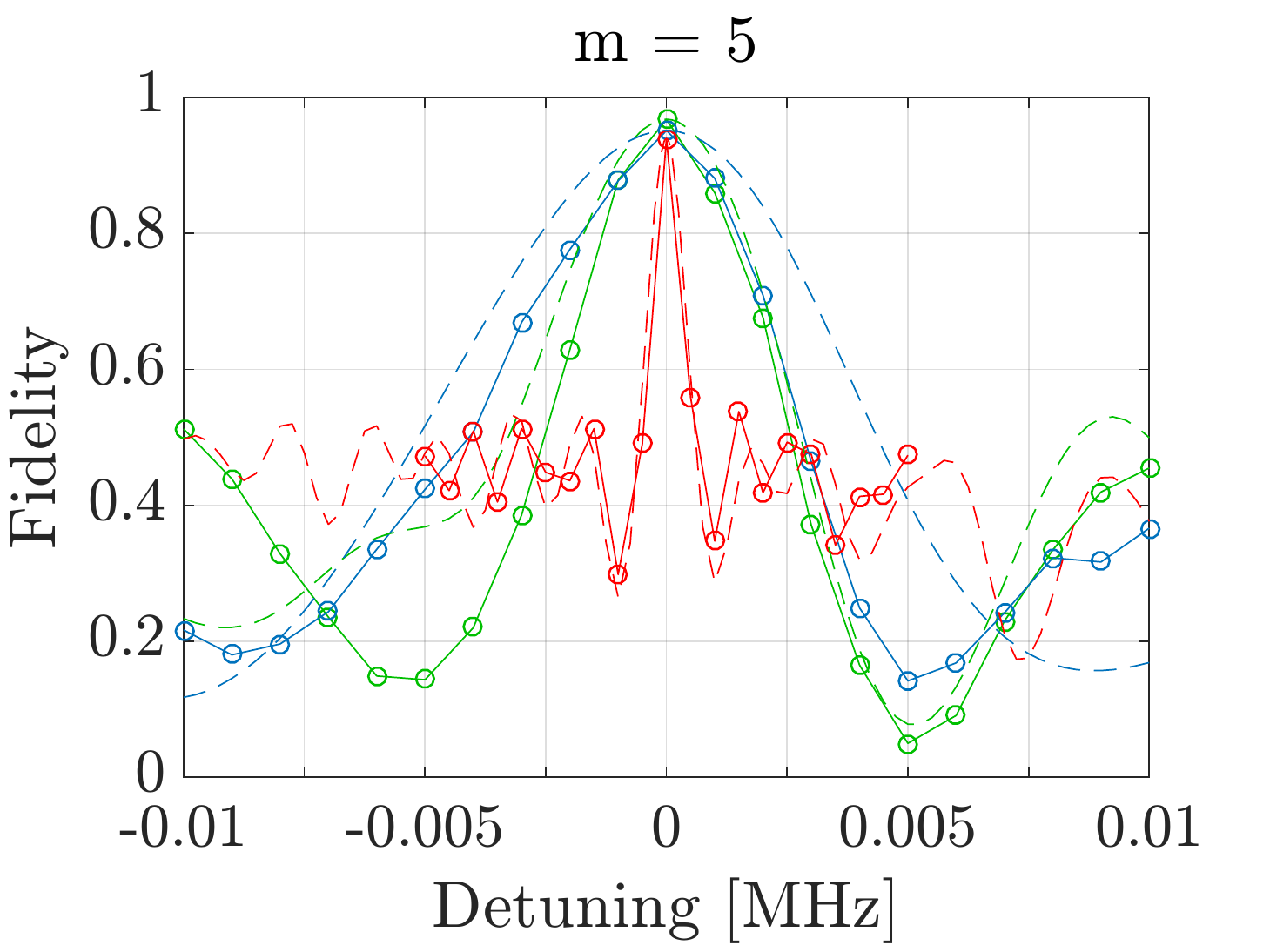} \label{5 pulses Fidelity vs detuning}}
\subfigure[]{\includegraphics[width=0.66 \columnwidth]{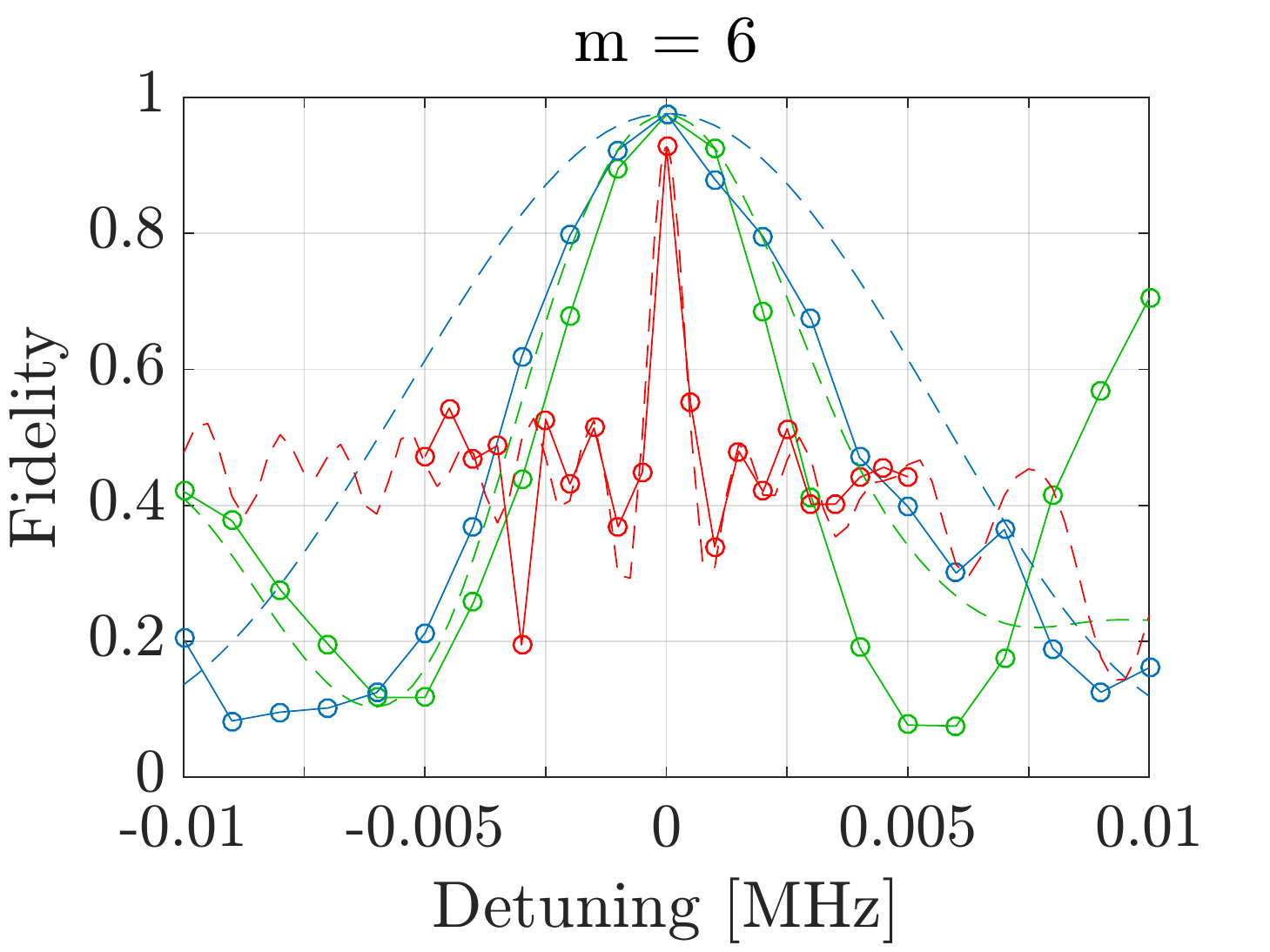} \label{6 pulses Fidelity vs detuning}}
\subfigure[]{\includegraphics[width=0.66 \columnwidth]{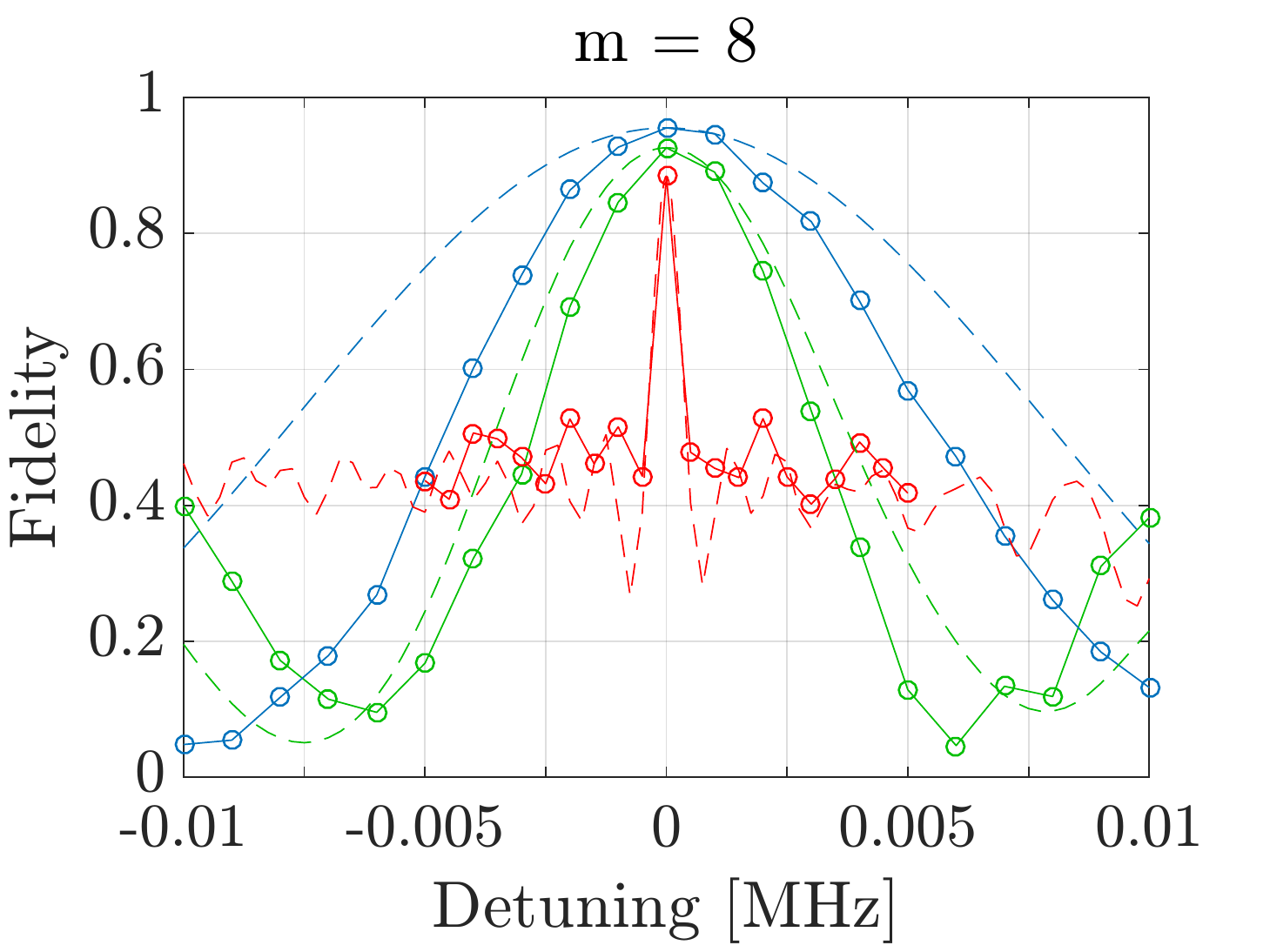} \label{8 pulses Fidelity vs detuning}}
\subfigure[]{\includegraphics[width=0.66 \columnwidth]{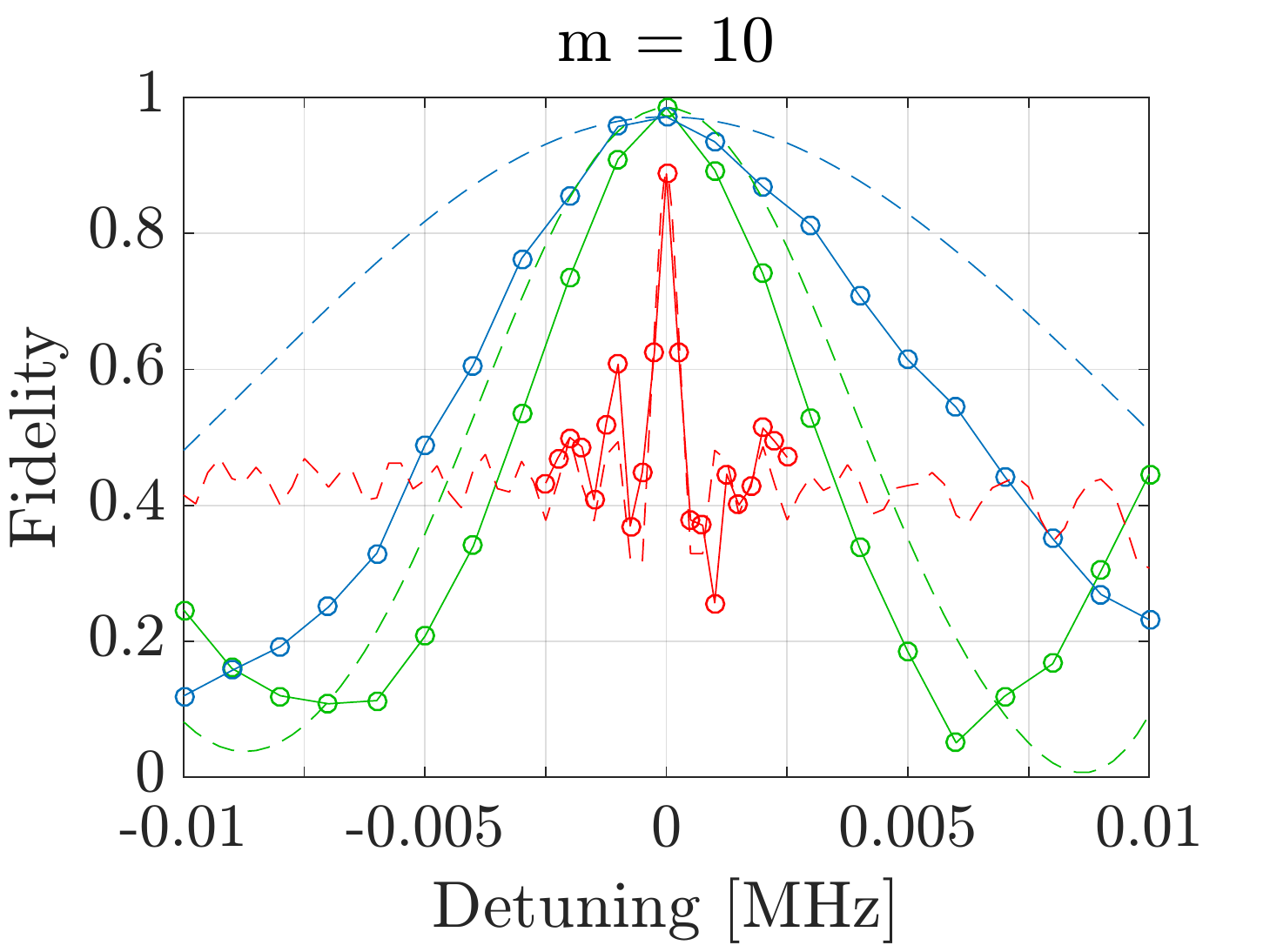} \label{10 pulses Fidelity vs detuning}}
\subfigure[]{\includegraphics[width=0.66 \columnwidth]{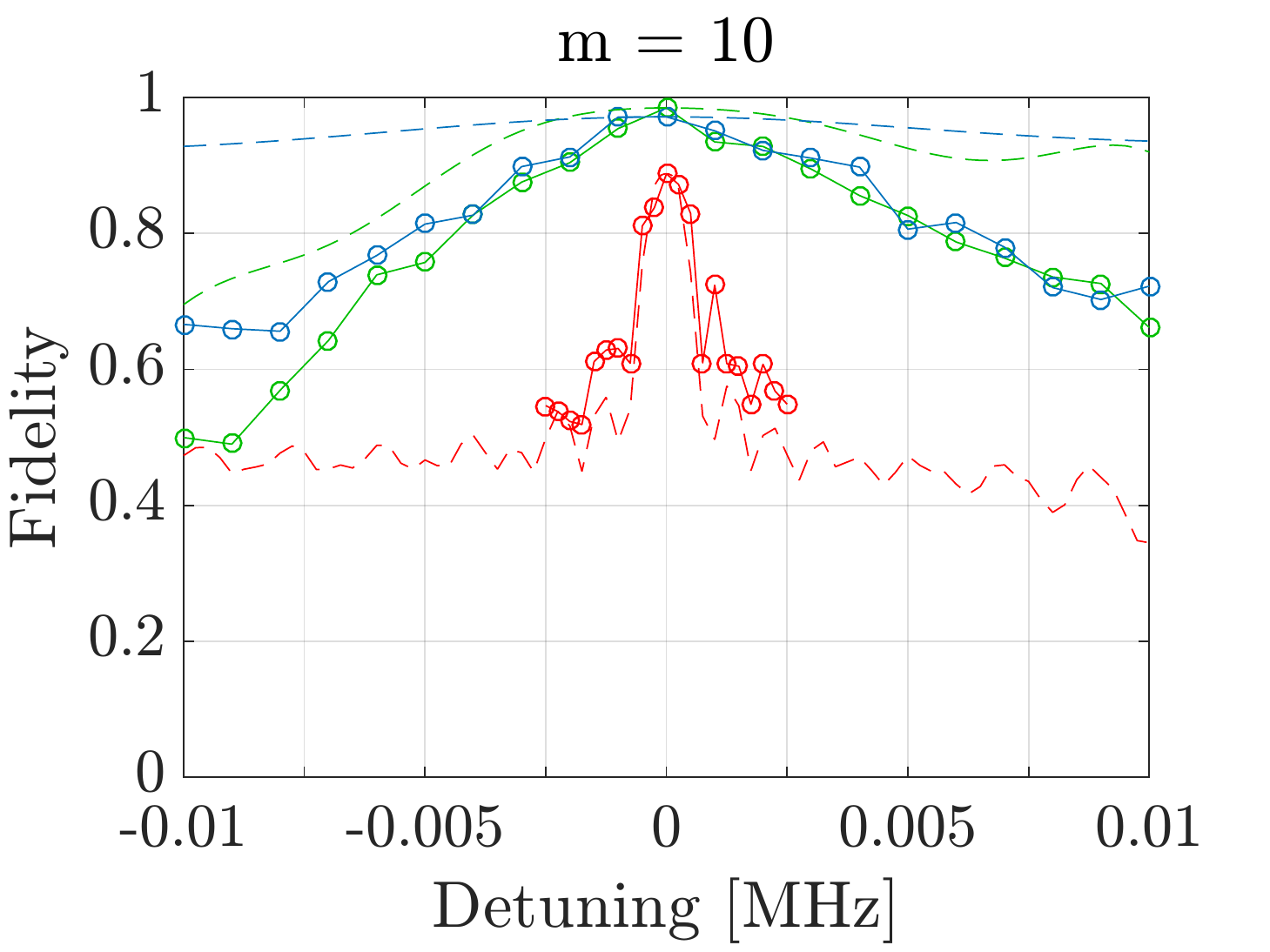} \label{10 pulses Fidelity vs detuning - Entanglement}}
\caption{An experimental and numerical comparison of $\sigma_{z}$ robustness for three  M{\o}lmer-S{\o}rensen entangling gate protocols: the fast DD scheme, as proposed in this article ({\color[rgb]{0 0.4470 0.7410} blue}); the slow DD scheme ({\color[rgb]{0 .75 0} green}); and the slow DD scheme without DD pulses ({\color{red} red}). The number of pulses is denoted as 'm'. The comparison is shown for differing pulse sequences enumerated by the number of DD arms. 
Two $^{88}Sr+$ were entangled according to the appropriate protocol using a 674 nm laser and the their final state measured via state-selective fluorescence. $\sigma_z$ noise was implemented by detuning all driving lasers from their resonance frequencies. The fidelity with the specific fully entangled state, calculated from measurement results, is shown. $95\%$ confidence intervals are under $\pm 0.03$ and are not plotted.  
The numerics were done for the MS Hamiltonian with the appropriate pulse sequence and a detuning term, alternating the dynamics between the MS Hamiltonian (Eq.~\ref{MS}) with detuning $\frac{\Delta}{2}\sum_i \sigma_{\beta,i}$, and the detuned $\pi$ pulse. The figures A-E shows the fidelity for a specific state, figure F shows the fidelity for an entangled state up to an arbitrary phase. F shows that high quality entangled states are achieved even at large detunings, albeit at a phase that differs from the zero detuning case.
}
\label{experiment}
\end{figure*}

We experimentally implement the fast dynamically decoupled gate with trapped ions and demonstrate its robustness to noise. Two $^{88}Sr+$  ions are spatially confined in a linear Paul trap with an axial frequency of $\nu=1.67$ MHz and radial frequencies of  $\sim4$ MHz \cite{Akerman2011}. A qubit is encoded on Zeeman-splitted sublevels of the $5S_{1/2}(m=-1/2)\rightarrow4D_{5/2}(m=1/2)$ optical electric-quadrupole transition of each ion with a natural lifetime of $\sim0.4$ seconds. This transition is driven using a narrow linewidth ($\sim 60$ Hz) $674$ nm laser locked to a stable Fabry-Perot cavity. State-selective fluorescence detection is performed by illuminating the ion with a  $422$ nm laser resonant with the $5S_{1/2}\rightarrow5P_{1/2}$ dipole allowed transition and collecting the fluorescence signal with an  EMCCD camera, enabling a non-ambivalent readout of the two ion state.  The ions are Doppler cooled, followed by sideband cooling of the center-of-mass motional component to the ground state. The M{\o}lmer-S{\o}rensen Hamiltonian  (Eq.~\ref{MS}) is enacted via bichromatic off-resonant  driving of the $5S_{1/2}\rightarrow4D_{5/2}$ with frequencies $\omega_{SD}\pm\left(\nu+\varepsilon\right)$, where $\omega_{SD}$ is the resonance carrier frequency \cite{Akerman2015}. Dynamical decoupling $\pi$ pulses are implemented by halting the bichromatic field operation and activating a monochromatic field with frequency $\omega_{SD}$.  In protocols where more than a single pulse is needed, the phases of consecutive pulses are flipped in order to reverse coherent build-up of error due to imperfect pulses. The typical coupling constants for the bichromatic and monochromatic fields are  $\eta\Omega_{B}\approx3$ kHz and $\Omega_{M}\approx170$ kHz, respectively. The bosonic mediator of interaction is the axial center-of-mass phononic mode.

In order to experimentally demonstrate the robustness of the dynamically-decoupled entangling gate to $\sigma_{z}$ type noise - which in ion traps is commonly a result of magnetic field  or laser phase fluctuations - we vary the detuning of the bichromatic and  monochromatic fields, perform the entangling gate protocol, and measure  the fidelity of the achieved state $\rho$ with respect to the desired fully entangled state $|\psi\rangle\equiv\frac{1}{\sqrt{2}}\left(|gg\rangle+i|ee\rangle\right)$  (Fig.~\ref{experiment}) . Varying the detuning adds a constant $\sigma_{z}$ term to the Hamiltonian similar to the effect of an external DC magnetic field.  The fidelity at the end of the gate is calculated as  $\mathcal{F}=\left\langle \psi\right|\rho\left|\psi\right\rangle $ which is the overlap squared of the measured state with the desired state.  We execute this experiment with 3 distinct protocols: (a) The fast dynamical  decoupling scheme, as detailed in the paragraphs above; (b) The slow scheme,  in which a $\pi$ pulse is applied only when the boson mediator is fully decoupled  from the qubit subspace; (c) The slow scheme without executing the DD pulses.  The latter acts as a reference to which one can compare the two DD schemes, thereby showing their meaningful impact. Furthermore, we execute all three protocols with different numbers of DD arms, ranging from four to ten. We compare the experimental results to a numerical simulation of the different protocols. 

The fast dynamically decoupled gate is shown to be more robust to $\sigma_z$ noise than its slow counterpart with the same number of DD pulses. Increasing the amount of pulses generates a marked robustness, particularly with the fast scheme; this is likely due to the significantly higher DD frequency of this protocol. Measuring final state fidelity with respect to some maximally entangled state at arbitrary phase shows that the generation of entanglement is fairly robust, and that a considerable portion of fidelity loss with respect to the spcific required state at finite detuning is due to a phase shift of the entangled state. Fidelity drops as the detuning approaches $\epsilon$, the point at which the sideband transition is resonantly addressed. The reason for discrepancy between simulations and experiment is not known. Note that these measurements simulate DC noise only; for AC noise, the benefits of the fast scheme should be even more pronounced.

\section{Discussion}

One interesting application of the fast DD gate is in microwave-based trapped ion platforms. To overcome the negligible microwave photon recoil, the spin-motion interaction can originate from a static magnetic field gradient. This gives rise to the MS Hamiltonian (Eq. \ref{MS}) $\tilde{\Omega} \sum_i \sigma_{z, i} \bla{b^\dagger e^{i \varepsilon t} +H.c.}$ where $ \tilde{\Omega}=\mu_B X_{i,n} dB_z/dz $ is the Rabi frequency, $\mu_B$ is the Bohr magneton, $ X_{i,n}$ is the standard deviation of the $n^{th}$ vibrational mode and the $i^{th}$ ion, $dB_z/dz$ represents the magnetic field gradient in the $z$ axis, and $\varepsilon= \nu_n$ is the vibrational frequency of mode $n$. As the microwave qubits have to be magnetic field dependent, they are also sensitive to the ambient magnetic field fluctuations. To compensate for this noise, pulsed DD has been considered in ref \cite{Wunderlich2013prl}, however, due to the very high detunings $\tilde{\Omega} \ll\varepsilon$, the gate was performed in the slow interacting regime, and thus resulted in a very modest fidelity. By utilizing the fast gate, the number of pulses and their duration could be adjusted such that the gate can be realized in the fast interacting regime. In comparison to current microwave-based entangling gates \cite{Gatis2015NJP,Weidt2016PRL}, the flower gate can be more than order of magnitude faster, having a similar duration as ref \cite{Itsik2015NJP}. 

Many quantum systems use a boson to mediate the interaction between different qubits. Dynamical decoupling techniques can be used in order to mitigate the damage of noise on these systems. The naive approach of applying the dynamical decoupling $\pi$ pulses, at times when the spins are disentangled from the boson,
increases the gate duration by a factor of $\sqrt{N}$.  To overcome this issue, we have proposed to apply dynamical decoupling pulses with a certain time separation, such that higher levels of the boson degrees of freedom are populated. In this way, the dynamical decoupling pulses not only suppress the main noise sources during the boson mediated interaction, but also considerably reduce the dynamical decoupling time overhead, increasing the gate robustness to other uncompensated noises.

\paragraph{Acknowledgments---}
We acknowledge the support of the Israel Science Foundation(grant no. 1500/13), the Marie Curie Career Integration Grant (CIG) IonQuanSense(321798), the Niedersachsen-Israeli Research Cooperation Program, the US Army Research Office under Contract W911NF-15-1-0250,  the Crown Photonics Center, the Minerva Foundation, iCore - Israeli Excellence Center Circle of Light, and the European Research Council (consolidator grant 616919-Ionology).



                                          

\section{Supplementary Material}

\paragraph{Inducing entanglement between trapped ions---}  
We consider two ions that are trapped in a linear trap, and interact due to Coulomb repulsion. The two ions system is diagonalized in the harmonic approximation, thus obtaining the vibrational normal modes: the center of mass mode, and the stretch mode. Each mode can be quantized where a vibrational phonon is introduced. Apart from the external degrees of freedom of vibration, the ions possess internal degrees of freedom. Two long lived internal energy levels compose the pseudo-spin or a qubit. 

The typical micrometer ion separation neutralizes the direct magnetic dipole - dipole interaction. Therefore, to induce entanglement between the spins, the vibrational phonon is used to mediate the spin-spin interaction. To this end, a spin-phonon coupling should first be induced. 
The spin-phonon coupling could be induced by lasers, by near-field microwave control or by a combined microwave and a magnetic gradient system.
Optical forces induce a sufficiently large recoil; meaning, the short wave-length radiation is coupled to the motion, and thus to the vibrational phonons. In the microwave based implementation, on the other hand, the long wave-length radiation induces a negligible recoil. Therefore, a magnetic field gradient is introduced to induce the spin-phonon interaction.



For simplicity we consider the laser based implementation, having only one relevant vibrational mode that is used as the mediator of the spin-spin interaction. Therefore, the Hamiltonian including the spins, the vibrational phonon and the optical driving field reads
\begin{align}
\begin{split}
& \mathcal{H}_{optic} = \nu b^\dagger b + \sum_{i=1,2} \frac{\omega_0}{2} \sigma_{z,i} + \Omega \sigma_{x,i} \cos \blb{\omega_d t + k x} \\ 
& = \nu b^\dagger b + \sum_{i=1,2} \frac{\omega_0}{2}\sigma_{z,i} +\Omega \sigma_{x,i} \cos \blb{\omega_d t + \eta \left(b^\dagger +b\right)},
\label{optic}
\end{split}
\end{align}
where $\sigma_{\alpha,i}$ are the Pauli matrices of the $i^{th}$ spin and in the $\alpha^{th}$ direction, $b^\dagger$, $b$ are the phonon creation and annihilation operators, $\omega_0$ is the bare energy gap of the spin, $\nu$ is the vibration secular frequency, $\Omega$ is the Rabi frequency, $\omega_d$ is the driving frequency, $k$ is the optical momentum, $x=x_0 \left(b^\dagger +b\right)$ is the position operator,  $\eta=k x_0 $ is the Lamb-Dicke parameter and $x_0$ is the ground state width. 

Moving to the interaction picture with respect to the spin energy gap and the phonon term (first and second terms in Eq.~\ref{optic}) we obtain
\begin{align}
\begin{split}
\mathcal{H}_{optic_I} & =
 \frac{\Omega}{2} \sum_{i=1,2}\sigma_{+,i} e^{i\bla{\omega_0 - \omega_d} t} e^{-i\eta \bla{b^\dagger e^{i\nu t} +b e^{-i\nu t}}}  +h.c \\
& = \frac{\Omega}{2} \sum_{i=1,2}\sigma_{+,i} e^{i\bla{\omega_0 - \omega_d} t} \mathcal{D}\blb{-i\eta e^{i\nu t}}  +h.c.,
\label{optic2}
\end{split}
\end{align}
after omitting the counter rotating term, which are negligible in the rotating wave approximation (RWA), and where the displacement operator is introduced $ \mathcal{D}\blb{\alpha}=\exp\blb{\alpha b^\dagger - h.c}$. By operating in the Lamb-Dicke regime, namely $\eta\sqrt{\left\langle n \right\rangle+1} \ll 1$, with $\left\langle n \right\rangle$ being the average number of phonons, the displacement exponent in Eq.~\ref{optic2} can be expanded to the first order, where the Hamiltonian reads
\begin{multline}
\mathcal{H}_{optic_I} =\\
 \frac{\Omega}{2} \sum_{i=1,2}\sigma_{+,i} e^{i\bla{\omega_0 - \omega_d} t} \bla{1-i\eta \bla{b^\dagger e^{i\nu t} +b e^{-i\nu t}}}  +h.c.
\label{optic3}
\end{multline}
This Hamiltonian is the main tool of performing entangling gates with trapped ions. By imposing the right detuning we can generate the interaction between the different spins, which is obtained in the second order of perturbation theory.

\begin{figure}
\def \svgwidth{1 \columnwidth}
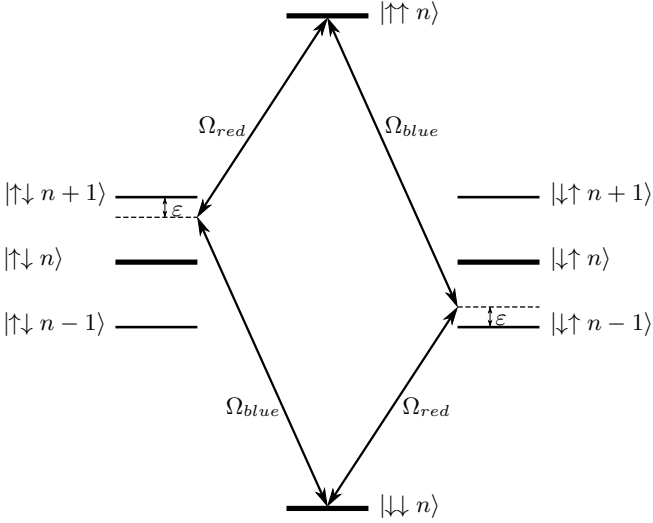
\caption{Energy scheme of two ions oscillating in collective vibrational mode. Been driven by two lasers that are detuned close to the red and blue sideband transition. Two paths for the $\ket{\hspace{-3px} \downarrow \downarrow n} \leftrightarrow \ket{\hspace{-3px} \uparrow \uparrow n}$ transition are indicated.} \label{mslevel}
\end{figure}

\paragraph{M{\o}lmer S{\o}rensen gate.---}
Imprinting an additional frequency onto the resonant laser beam gives rise to two beatnotes that are equally detuned from the carrier transition $\omega_d - \omega_0=\delta_\pm = \pm\bla{\nu- \varepsilon}$. The positive detuning $\delta_+$ generates the $-\varepsilon$ detuned blue sideband transition out from Eq.~\ref{optic3}, whereas the negative detuning generates the $\varepsilon$ detuned blue sideband transition (Fig.~\ref{mslevel}):
\bea
\mathcal{H}_{blue} &=& -\frac{i\eta\Omega}{2} \sum_{i=1,2}\sigma_{+,i} b^\dagger e^{i\varepsilon t}   +h.c, \label{blue}\\
\mathcal{H}_{red} &=& -\frac{i\eta\Omega}{2} \sum_{i=1,2}\sigma_{+,i} b e^{-i\varepsilon t}  +h.c,
\label{red}
\eea
after omitting the fast rotating terms whose contribution is negligible in the RWA, assuming the following parameter hierarchy  $\Omega\eta \sim \varepsilon \ll \Omega \ll \nu$. Therefore, combining the two beatnotes together (Eq.~\ref{blue},\ref{red}) results in the MS Hamiltonian:
\beq
\mathcal{H}_{MS} = \frac{\eta\Omega}{2} \sum_{i=1,2}\sigma_{y,i} \bla{b^\dagger e^{i\varepsilon t} +b e^{-i\varepsilon t}},
\label{MS_I}
\eeq
which is Eq.~1 in the main part with $\tilde{\Omega}=\eta\Omega/2$. 

\paragraph{Single sideband gate.---} Another approach to generating the spin-phonon interaction, and obtaining the MS Hamiltonian (Eq.~1)  is the single sideband scheme \cite{Bermudez2012PRA,Tan2013PRL}. Using two driving fields, the first one for driving the red sideband transition (Eq.~\ref{red}) and the second one for driving the carrier transition, by setting $\omega_d=\omega_0$ in Eq.~\ref{optic3}, the Hamiltonian now reads
\beq
\mathcal{H}_I = \frac{\Omega_c}{2} \sigma_x - i \frac{\Omega \eta}{4} \left( \left( \sigma_x + i \sigma_y \right) b^\dagger e^{i \varepsilon t} + H.c \right). \label{bermudez}
\eeq
By moving to the rotating frame of the carrier transition, $\Omega_c \sigma_x / 2$ the single sideband Hamiltonian becomes
\beq
\mathcal{H}_{SM} =\frac{\Omega \eta}{4 i} \sigma_x b^\dagger e^{i \varepsilon t} + H.c \label{bermudezRWA}
\eeq
after omitting the fast oscillating $\sigma_y$ terms in the RWA, assuming $\varepsilon \ll \Omega_c$.
Thus, Eq.~\ref{bermudezRWA}, is the MS Hamiltonian (Eq.~1) obtained for a different basis, with $\tilde{\Omega} = \Omega \eta / 4i$. Note that regarding the other quantum architectures, the natural spin-boson coupling is the Jaynes��-Cummings interaction,  which is equivalent to the ion's red sideband transition. It is challenging to generate the anti-Jaynes-Cummings interaction in addition; therefore, the way to generate the MS gate is the single sideband approach.

\paragraph{Fidelity for the MS gate with detuning $\Delta$ for large number of pulses ---}

The fidelity can be approximated, in the limit of many DD pulses, by the errors accumulated for a single part of the gate (i.e. MS interaction for time $\delta t$, followed by DD pulse and another MS interaction for the same time $\delta t$).
\begin{align}
F \approx \frac{1}{4} \bra{\alpha} Tr_{spin}\blb{\bla{U\bla{t_0+\delta t,t_0} U_{opt}^\dagger\bla{t_0+\delta t,t_0}}^N} \ket{\alpha}
\end{align}
where $U\bla{t_0+\delta t,t_0}$ is the unitary evolution of the MS Hamiltonian with a detuning term $\frac{\Delta}{2}\sigma_z$, and $U_{opt}\bla{t_0+\delta t,t_0}$ is the same evolution without detuning. The fidelity is most damaged by the unwanted displacement that is caused by the rotated MS interaction, This can be seen in the second order Magnus expansion of the detuned Hamiltonian. The infidelity is then given by the amount of displacement,
\begin{align}
IF\approx & \left(\frac{\Delta}{2\varepsilon}\frac{\Omega\eta}{2\varepsilon}N\right)^{2}\left|g\left(\varepsilon \delta t\right)\right|^{2}\\
g\left(\varepsilon \delta t\right)= & \left(\left(\frac{1}{2}-i\right)\varepsilon \delta t+\frac{\left(\varepsilon \delta t\right)^{2}}{2}\right) + \\ \nonumber 
&+\left(i+\left(-\frac{5}{2}+i2\right)\varepsilon \delta t\right)e^{i\varepsilon \delta t}+\\ \nonumber 
&+\left(-i+\left(1-i\right)\varepsilon \delta t-\frac{\left(\varepsilon \delta t\right)^{2}}{2}\right)e^{i2\varepsilon \delta t} \nonumber 
\end{align}

where $g\bla{\varepsilon \delta t}$ is found using the 2nd order Magnus expansion.

For both the flower sequence, and the echo sequence, the $g\bla{\varepsilon \delta t}$ is weakly dependent on the number of pulses (or $\varepsilon \delta t$), and equals roughly $5.2$ for the flower, and $1.6$ for the echo. The main difference comes from $\varepsilon$ it self, as for the flower $\varepsilon \sim N$ for $N\gg 1$, thus $IF \sim N^{-2}$, and for the echo sequence $\varepsilon \sim \sqrt{N}$, thus $IF\sim const$. For finite pulse time, when the total pulse time ($=N\times$(pulse time)) is comparable with the total gate time, both scaling will go as $IF\sim \bla{\Delta \tau N}^2$ due to the prolong gate time.

{\em XY8 pulse sequence ---} The sequence that we have presented utilizes pulses in a direction which is orthogonal to $\alpha$. Pulses in the $\alpha$ direction can be added trivially as these do not change the sequence of the gate. This gives the ability to realize XY4 and XY8 pulses. 

The XY4 sequence could be realized as it is as the extra $\sigma_x$ operation commutes with the gate operators and thus does not change the structure of the gate; meaning the path in phase space stays exactly as realized in this work. The XY4 sequence is however is not robust as the XY8\cite{Farfurnik,Wang2012,Wang2012a}.

\begin{figure}
\includegraphics[width=1 \columnwidth]{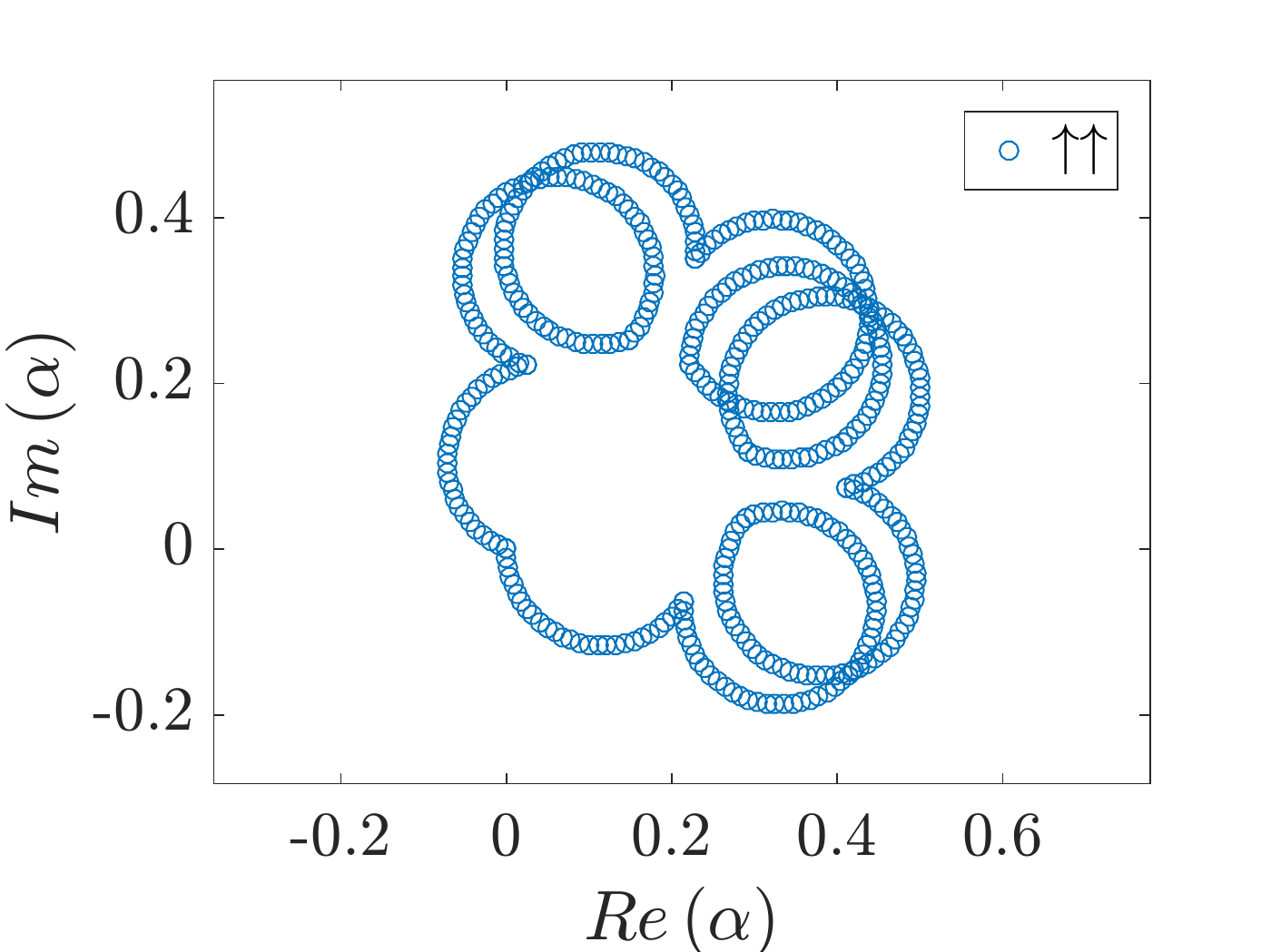} \label{XY8 traj}
\caption{Numerical simulation of the phase space diagram showing the dynamics of a state starting at $\ket{\uparrow\uparrow}$ under the MS Hamiltonian (Eq.~\ref{MS_I}) and an XY8 pulse sequence. Simulated with finite pulse time, where the center of the pulses are at times $\varepsilon t \approx \bla{1,5,11,15}\cdot 1.78$. The detuning was adjust numerically to give the CNOT gate, and is given by $\varepsilon \approx \Omega \eta \cdot 6.07.$
} \label{xy8}
\end{figure}

An XY8 version could also be constructed in the following way.
For a sequence of pulses at times $\left\{ t_j \right\}_{j=1}^n$ the overall displacement is given by the sum of all the MS SDK 
\begin{multline}
U_{MS}\left(t\right) = \mathcal{D}\blb{\frac{\tilde{\Omega}}{\varepsilon}\sum_{i=1,2}\sigma_{\alpha,i}\left(1-e^{i\varepsilon t}\right)} \cdot\\
\exp \blb{ i\bla{\frac{\tilde{\Omega}}{\varepsilon}\sum_{i=1,2}\sigma_{\alpha,i}  }^{2}\bla{\varepsilon t-\sin\bla{\varepsilon t}} },
\label{unitary}
\end{multline}
 and thus proportional to $ dis =  \sum_{j=1}^n \bla{-1}^j \bla{e^{i \varepsilon t_j} - e^{i \varepsilon t_{j-1}}}$.
 For finite pulse times one must adjust for the time delay when the MS interaction is off and the trap keeps rotating, thus we get
\begin{eqnarray}
dis  &= & \bla{e^{i \varepsilon \bla{t_1-\tau/2}} - 1} + \nonumber \\ 
&+ & \sum_{j=1}^{n-1} \bla{-1}^j \bla{e^{i \varepsilon \bla{t_j-\tau/2}} - e^{i \varepsilon \bla{t_{j-1}+\tau/2}}}  +\nonumber  \\ 
&+& \bla{-1}^n \bla{e^{i \varepsilon t_j} - e^{i \varepsilon \bla{t_{j-1}+\tau/2}}}
\end{eqnarray}
By setting the overall displacement to zero, we make sure that the spins are disentangled from the phonons at the end of the gate. For an XY8 sequence this can be easily done numerically. For instance, for the XY8 sequence (Fig.~\ref{xy8}) the pulses are applied at times $\varepsilon t \approx \bla{1,5,11,15}\cdot 1.78$, and the detuning is given by $\varepsilon \approx \Omega \eta \cdot 6.07$. Where we ignored the pulses that commute with the MS interaction as these do not create the effective SDK behavior.


\begin{figure*} 
\subfigure[]{\includegraphics[width=0.66 \columnwidth]{"F_vs_detuning_m=4"} \label{4 pulses Fidelity vs detuning}}
\subfigure[]{\includegraphics[width=0.66 \columnwidth]{"F_vs_detuning_m=5"} \label{5 pulses Fidelity vs detuning}}
\subfigure[]{\includegraphics[width=0.66 \columnwidth]{"F_vs_detuning_m=6"} \label{6 pulses Fidelity vs detuning}}
\subfigure[]{\includegraphics[width=0.66 \columnwidth]{"F_vs_detuning_m=8"} \label{8 pulses Fidelity vs detuning}}
\subfigure[]{\includegraphics[width=0.66 \columnwidth]{"F_vs_detuning_m=10"} \label{10 pulses Fidelity vs detuning}}
\subfigure[]{\includegraphics[width=0.66 \columnwidth]{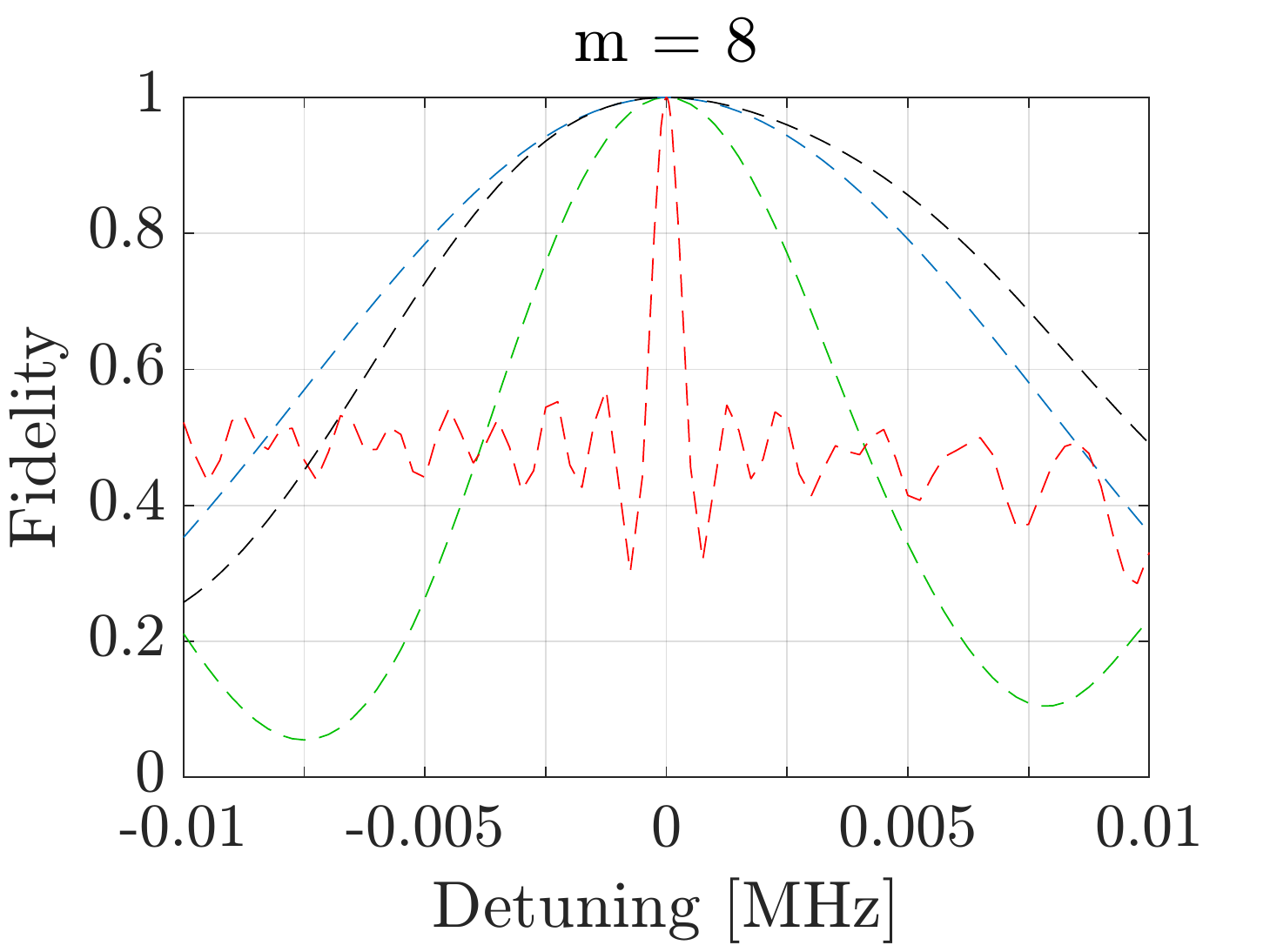} \label{XY8 Fidelity}}
\caption{These figures show a comparison of three schemes, the Flower scheme ({\color[rgb]{0 0.4470 0.7410} blue}), the Echo scheme ({\color[rgb]{0 .75 0} green}), and without pulses ({\color{red} red}) (same detuning as in the Echo scheme).
The numerics was done for the MS Hamiltonian with the appropriate pulse sequence and a detuning term, alternating the dynamics between the MS Hamiltonian (Eq.~\ref{MS_I}) with detuning $\frac{\Delta}{2}\sum_i \sigma_{\beta,i}$, and the detuned $\pi$ pulse.
}
\end{figure*}



\begin{thebibliography}{0}%
\makeatletter
\providecommand \@ifxundefined [1]{%
 \@ifx{#1\undefined}
}%
\providecommand \@ifnum [1]{%
 \ifnum #1\expandafter \@firstoftwo
 \else \expandafter \@secondoftwo
 \fi
}%
\providecommand \@ifx [1]{%
 \ifx #1\expandafter \@firstoftwo
 \else \expandafter \@secondoftwo
 \fi
}%
\providecommand \natexlab [1]{#1}%
\providecommand \enquote  [1]{``#1''}%
\providecommand \bibnamefont  [1]{#1}%
\providecommand \bibfnamefont [1]{#1}%
\providecommand \citenamefont [1]{#1}%
\providecommand \href@noop [0]{\@secondoftwo}%
\providecommand \href [0]{\begingroup \@sanitize@url \@href}%
\providecommand \@href[1]{\@@startlink{#1}\@@href}%
\providecommand \@@href[1]{\endgroup#1\@@endlink}%
\providecommand \@sanitize@url [0]{\catcode `\\12\catcode `\$12\catcode
  `\&12\catcode `\#12\catcode `\^12\catcode `\_12\catcode `\%12\relax}%
\providecommand \@@startlink[1]{}%
\providecommand \@@endlink[0]{}%
\providecommand \url  [0]{\begingroup\@sanitize@url \@url }%
\providecommand \@url [1]{\endgroup\@href {#1}{\urlprefix }}%
\providecommand \urlprefix  [0]{URL }%
\providecommand \Eprint [0]{\href }%
\providecommand \doibase [0]{http://dx.doi.org/}%
\providecommand \selectlanguage [0]{\@gobble}%
\providecommand \bibinfo  [0]{\@secondoftwo}%
\providecommand \bibfield  [0]{\@secondoftwo}%
\providecommand \translation [1]{[#1]}%
\providecommand \BibitemOpen [0]{}%
\providecommand \bibitemStop [0]{}%
\providecommand \bibitemNoStop [0]{.\EOS\space}%
\providecommand \EOS [0]{\spacefactor3000\relax}%
\providecommand \BibitemShut  [1]{\csname bibitem#1\endcsname}%
\let\auto@bib@innerbib\@empty
\end{thebibliography}%


\begin{references}

\bibitem{Molmer1999prl} A. S{\o}rensen and K. M{\o}lmer \href{http://journals.aps.org/prl/abstract/10.1103/PhysRevLett.82.1971}{Phys. Rev. Lett. {\bf 82}, 1971 (1999).} 

\bibitem{Molmer2000pra}  K. M{\o}lmer and A. S{\o}rensen \href{http://journals.aps.org/pra/abstract/10.1103/PhysRevA.62.022311}{Phys. Rev. A {\bf 62}, 022311 (2000).}

\bibitem{Milburn} GJ Milburn, S Schneider, DFV James, \href{http://iontrap.umd.edu/wp-content/uploads/2013/10/Ion-trap-quantum-computing-with-warm-ions.pdf}{Fortschritte der Physik {\bf 48}, 801  (2000)}

\bibitem{Wineland2000N} C. A. Sackett, D. Kielpinski, B. E. King, C. Langer, V. Meyer, C. J. Myatt, M. Rowe, Q. A. Turchette, W. M. Itano, D. J. Wineland, and C. Monroe, \href{http://www.nature.com/nature/journal/v404/n6775/full/404256a0.html}{Nature {\bf 404}, 256-259 (2000).}

\bibitem{Blatt2014NPH} J. Benhelm, G. Kirchmair, C. F. Roos, and R. Blatt, \href{http://www.nature.com/nphys/journal/v4/n6/full/nphys961.html}{Nature Physics {\bf 4}, 463 - 466 (2008).}

\bibitem{Monroe2009PRL} K. Kim, M.-S. Chang, R. Islam, S. Korenblit, L.-M. Duan, and C. Monroe \href{http://journals.aps.org/prl/abstract/10.1103/PhysRevLett.103.120502}{Phys. Rev. Lett. {\bf 103}, 120502  (2009).}

\bibitem{Ozeri2014PRA} N. Navon, N. Akerman, S. Kotler, Y. Glickman, and R. Ozeri \href{http://journals.aps.org/pra/abstract/10.1103/PhysRevA.90.010103}{Phys. Rev. A {\bf 90}, 010103(R) (2014).}

\bibitem{Tan2} T. R. Tan,	J. P. Gaebler,	Y. Lin,	Y. Wan,	R. Bowler,	D. Leibfried, and D. J. Wineland, \href{http://www.nature.com/nature/journal/v528/n7582/full/nature16186.html}{Nature {\bf 528}, 380–383 (2015).}

\bibitem{Didi2003nature} D. Leibfried, B. DeMarco, V. Meyer, D. Lucas, M. Barrett, J. Britton, W. M. Itano, B. Jelenkovic acute, C. Langer, T. Rosenband, and D. J. Wineland \href{http://www.nature.com/nature/journal/v422/n6930/full/nature01492.html}{Nature {\bf 422}, 412-415 (2003).}

\bibitem{Lucas2015Nature1} C. J. Ballance,	V. M. Schafer,	J. P. Home,	D. J. Szwer,	S. C. Webster,	D. T. C. Allcock,	N. M. Linke,	T. P. Harty,	D. P. L. Aude Craik,	D. N. Stacey,	A. M. Steane	and D. M. Lucas, \href{http://www.nature.com/nature/journal/v528/n7582/full/nature16184.html}{Nature {\bf 528}, 384–386 (2015).}

\bibitem{Lucas2016prl} C. J. Ballance, T. P. Harty, N. M. Linke, M. A. Sepiol, D. M. Lucas, \href{https://journals.aps.org/prl/abstract/10.1103/PhysRevLett.117.060504}{Phys. Rev. Lett. 117, 060504 (2016).}

\bibitem{harty2016prl} 
T. P. Harty et. al., \href{https://journals.aps.org/prl/abstract/10.1103/PhysRevLett.117.140501}{Phys. Rev. Lett. {\bf 117,} 140501 (2016).}

\bibitem{Bermudez2012PRA} A. Bermudez, P. O. Schmidt, M. B. Plenio, and A. Retzker  \href{http://journals.aps.org/pra/abstract/10.1103/PhysRevA.85.040302} {Phys. Rev. A 85, 040302(R) (2012).}

\bibitem{Itsik2015NJP} I. Cohen, S. Weidt, W. K. Hensinger, A. and Retzker, \href{http://iopscience.iop.org/article/10.1088/1367-2630/17/4/043008}{New J. Phys. {\bf 17}, 043008 (2015).}

\bibitem{Tan2013PRL} T. R. Tan, J. P. Gaebler, R. Bowler, Y. Lin, J. D. Jost, D. Leibfried, and D. J. Wineland, \href{http://journals.aps.org/prl/abstract/10.1103/PhysRevLett.110.263002}{Phys. Rev. Lett. {\bf 110}, 263002 (2013).}



\bibitem{Zhu2005prl} S-L. Zhu, Z. D. Wang, and P. Zanardi \href{http://journals.aps.org/prl/abstract/10.1103/PhysRevLett.94.100502}{Phys. Rev. Lett. {\bf 94}, 100502 (2005).}


\bibitem{Gao} Shi-Biao Zheng and Guang-Can Guo,\href{https://journals.aps.org/prl/abstract/10.1103/PhysRevLett.85.2392}{Phys. Rev. Lett. {\bf 85,} 2392 (2000)}

\bibitem{Majer2007nature} J. Majer, J. M. Chow, J. M. Gambetta, Jens Koch, B. R. Johnson, J. A. Schreier, L. Frunzio, D. I. Schuster, A. A. Houck, A. Wallraff, A. Blais, M. H. Devoret, S. M. Girvin, and R. J. Schoelkopf \href{http://www.nature.com/nature/journal/v449/n7161/full/nature06184.html}{Nature {\bf 449}, 443-447 (2007).}

\bibitem{Bennett2013prl} S. D. Bennett, N. Y. Yao, J. Otterbach, P. Zoller, P. Rabl, and M. D. Lukin \href{http://journals.aps.org/prl/abstract/10.1103/PhysRevLett.110.156402}{Phys. Rev. Lett. {\bf 110}, 156402 (2013).}

\bibitem{Alex2013njp} A. Albrecht, A. Retzker, F. Jelezko, and M. B. Plenio \href{http://iopscience.iop.org/article/10.1088/1367-2630/15/8/083014/meta}{New Journal of Physics {\bf 15}, 083014 (2013). }

%

\bibitem{Duan}
L.-M. Duan, M. D. Lukin, J. I. Cirac and P. Zoller, \href{http://www.nature.com/nature/journal/v414/n6862/abs/414413a0.html}{Nature {\bf 414,} 413 (2001)}

\bibitem{Mintert}
F. Mintert and C. Wunderlich
\href{https://journals.aps.org/prl/abstract/10.1103/PhysRevLett.87.257904}{ Phys. Rev. Lett. {\bf 87}, 257904  (2001)}

\bibitem{Chou} Chou et. al.,
\href{http://science.sciencemag.org/content/316/5829/1316}{Science
{\bf 316,} 1316 (2007)}

\bibitem{Solano2003prl} E. Solano, G. S. Agarwal, and H. Walther \href{http://journals.aps.org/prl/abstract/10.1103/PhysRevLett.90.027903}{Phys. Rev. Lett. {\bf 90}, 027903 (2003).}
 
\bibitem{Roos1} G. Kirchmair, J. Benhelm, F. Z\"{a}hringer, R. Gerritsma, C. F. Roos and R. Blatt \href{http://iopscience.iop.org/article/10.1088/1367-2630/11/2/023002?fromSearchPage=true}{New Journal of Physics, Volume {\bf 11}, 023002 (2009).}

\bibitem{Hanh1950} E. L. Hahn \href{http://journals.aps.org/pr/abstract/10.1103/PhysRev.80.580}{Phys. Rev. {\bf 80}, 580 (1950).} 

\bibitem{Viola1998pra} L. Viola and S. Lloyd \href{http://journals.aps.org/pra/abstract/10.1103/PhysRevA.58.2733}{Phys. Rev. A {\bf 58}, 2733 (1998).} 




\bibitem{DEER3} de Lange, G., van der Sar, T., Blok, M., Wang, Z-H.,
Dobrovitski, V. and Hanson, R. Sci. Rep. 2, (2012).





\bibitem{Hayes2012prl} D. Hayes, S. M. Clark, S. Debnath, D. Hucul, I. V. Inlek, K. W. Lee, Q. Quraishi, and C. Monroe \href{http://link.aps.org/doi/10.1103/PhysRevLett.109.020503}{Phys. Rev. Lett. {\bf 109}, 020503 (2012).}

\bibitem{Green2015prl} T. J. Green and M. J. Biercuk \href{http://link.aps.org/doi/10.1103/PhysRevLett.114.120502}{Phys. Rev. Lett. {\bf 114}, 120502 (2015).}

\bibitem{Ivanov2015pra} Svetoslav S. Ivanov and Nikolay V. Vitanov \href{http://link.aps.org/doi/10.1103/PhysRevA.92.022333}{Phys. Rev. A {\bf 92}, 022333 (2015).}

\bibitem{Cai2012njp} J-M. Cai, B Naydenov, R. Pfeiffer, L. P. McGuinness, K. D. Jahnke, F. Jelezko, M. B. Plenio, and A. Retzker \href{http://iopscience.iop.org/article/10.1088/1367-2630/14/11/113023/meta}{New Journal of Physics {\bf 14}, 113023 (2012).}

\bibitem{DEER1} T. Yamamoto. et al., \href{http://journals.aps.org/prb/abstract/10.1103/PhysRevB.88.201201}{Phys. Rev. B {\bf 88}, 201201(R) (2013)}

\bibitem{DEER2} B. Grotz, et. al., \href{http://iopscience.iop.org/article/10.1088/1367-2630/13/5/055004/meta}{New J. Phys. {\bf 13}, 055004 (2011)}

\bibitem{Garcia2003prl} J. J. Garc\'{\i}a-Ripoll, P. Zoller and J. I. Cirac \href{http://link.aps.org/doi/10.1103/PhysRevLett.91.157901}{Phys. Rev. Lett. {\bf 91}, 157901 (2003).}

\bibitem{Mizrahi2014apb} J. Mizrahi, B. Neyenhuis, K. Johnson, W. C. Campbell, C. Senko, D. Hayes, C. Monroe \href{http://dx.doi.org/10.1007/s00340-013-5717-6}{J. Appl. Phys. B {\bf 114}, 1, 45 (2014).}



\bibitem{Akerman2011}
N. Akerman et. al., \href{https://link.springer.com/article/10.1007%2Fs00340-011-4807-6}{ Applied Physics B, 107(4):1167--1174, ( 2011).}

\bibitem{Akerman2015}
N. Akerman et. al., \href{http://iopscience.iop.org/article/10.1088/1367-2630/17/11/113060/meta}{New Journal of Physics, 17(11):113060, (2015).}


\bibitem{Farfurnik}
D. Farfurnik et. al., \href{https://journals.aps.org/prb/abstract/10.1103/PhysRevB.92.060301}{Phys. Rev. B 92, 060301(R) (2015).}

\bibitem{Wang2012}
Zhi-Hui Wang et. al., \href{https://journals.aps.org/prb/abstract/10.1103/PhysRevB.85.085206}{Phys. Rev. B 85, 085206  (2012).}

\bibitem{Wunderlich2013prl}
Ch. Piltz, B. Scharfenberger, A. Khromova, A. F. Varón, and Ch. Wunderlich, \href{https://doi.org/10.1103/PhysRevLett.110.200501}{Phys. Rev. Lett. 110, 200501 (2013).}

\bibitem{Gatis2015NJP}
G. Mikelsons, I. Cohen, A. Retzker, and M. B. Plenio \href{http://iopscience.iop.org/article/10.1088/1367-2630/17/5/053032/pdf}{New J. Phys. {\bf 17}, 053032 (2015).}
\bibitem{Weidt2016PRL}
S. Weidt, J. Randall, S. C. Webster, K. Lake, A. E. Webb, I. Cohen, T. Navickas, B. Lekitsch, A. Retzker, and W. K. Hensinger \href{https://journals.aps.org/prl/abstract/10.1103/PhysRevLett.117.220501}
{Phys. Rev. Lett. {\bf 117,} 220501 (2016).}


\end{references}

\begin{references}

\bibitem{Molmer1999prl} A. S{\o}rensen and K. M{\o}lmer \href{http://journals.aps.org/prl/abstract/10.1103/PhysRevLett.82.1971}{Phys. Rev. Lett. {\bf 82}, 1971 (1999).} 

\bibitem{Molmer2000pra}  K. M{\o}lmer and A. S{\o}rensen \href{http://journals.aps.org/pra/abstract/10.1103/PhysRevA.62.022311}{Phys. Rev. A {\bf 62}, 022311 (2000).}

\bibitem{Wineland2000N} C. A. Sackett, D. Kielpinski, B. E. King, C. Langer, V. Meyer, C. J. Myatt, M. Rowe, Q. A. Turchette, W. M. Itano, D. J. Wineland, and C. Monroe, \href{http://www.nature.com/nature/journal/v404/n6775/full/404256a0.html}{Nature {\bf 404}, 256-259 (2000).}

\bibitem{Blatt2014NPH} J. Benhelm, G. Kirchmair, C. F. Roos, and R. Blatt, \href{http://www.nature.com/nphys/journal/v4/n6/full/nphys961.html}{Nature Physics {\bf 4}, 463 - 466 (2008).}

\bibitem{Monroe2009PRL} K. Kim, M.-S. Chang, R. Islam, S. Korenblit, L.-M. Duan, and C. Monroe \href{http://journals.aps.org/prl/abstract/10.1103/PhysRevLett.103.120502}{Phys. Rev. Lett. {\bf 103}, 120502  (2009).}

\bibitem{Ozeri2014PRA} N. Navon, N. Akerman, S. Kotler, Y. Glickman, and R. Ozeri \href{http://journals.aps.org/pra/abstract/10.1103/PhysRevA.90.010103}{Phys. Rev. A {\bf 90}, 010103(R) (2014).}

\bibitem{Tan2} T. R. Tan,	J. P. Gaebler,	Y. Lin,	Y. Wan,	R. Bowler,	D. Leibfried, and D. J. Wineland, \href{http://www.nature.com/nature/journal/v528/n7582/full/nature16186.html}{Nature {\bf 528}, 380–383 (2015).}

\bibitem{Didi2003nature} D. Leibfried, B. DeMarco, V. Meyer, D. Lucas, M. Barrett, J. Britton, W. M. Itano, B. Jelenkovic acute, C. Langer, T. Rosenband, and D. J. Wineland \href{http://www.nature.com/nature/journal/v422/n6930/full/nature01492.html}{Nature {\bf 422}, 412-415 (2003).}

\bibitem{Lucas2015Nature1} C. J. Ballance,	V. M. Schafer,	J. P. Home,	D. J. Szwer,	S. C. Webster,	D. T. C. Allcock,	N. M. Linke,	T. P. Harty,	D. P. L. Aude Craik,	D. N. Stacey,	A. M. Steane	and D. M. Lucas, \href{http://www.nature.com/nature/journal/v528/n7582/full/nature16184.html}{Nature {\bf 528}, 384–386 (2015).}

\bibitem{Lucas2015arXiv} C. J. Ballance, T. P. Harty, N. M. Linke, M. A. Sepiol, D. M. Lucas, \href{http://arxiv.org/abs/1512.04600}{arXiv:1512.04600 (2015).}

\bibitem{Bermudez2012PRA} A. Bermudez, P. O. Schmidt, M. B. Plenio, and A. Retzker  \href{http://journals.aps.org/pra/abstract/10.1103/PhysRevA.85.040302} {Phys. Rev. A 85, 040302(R) (2012).}

\bibitem{Itsik2015NJP} I. Cohen, S. Weidt, W. K. Hensinger, A. and Retzker, \href{http://iopscience.iop.org/article/10.1088/1367-2630/17/4/043008}{New J. Phys. {\bf 17}, 043008 (2015).}

\bibitem{Tan2013PRL} T. R. Tan, J. P. Gaebler, R. Bowler, Y. Lin, J. D. Jost, D. Leibfried, and D. J. Wineland, \href{http://journals.aps.org/prl/abstract/10.1103/PhysRevLett.110.263002}{Phys. Rev. Lett. {\bf 110}, 263002 (2013).}


\bibitem{Farfurnik}
D. Farfurnik et. al., \href{https://journals.aps.org/prb/abstract/10.1103/PhysRevB.92.060301}{Phys. Rev. B 92, 060301(R) (2015).}

\bibitem{Wang2012}
Zhi-Hui Wang et. al., \href{https://journals.aps.org/prb/abstract/10.1103/PhysRevB.85.085206}{Phys. Rev. B 85, 085206  (2012).}

\bibitem{Wang2012a}
Zhi-Hui Wang et. al., \href{https://journals.aps.org/prb/abstract/10.1103/PhysRevB.85.155204}{Phys. Rev. B 85, 155204  (2012).}



\end{references}
\end{document}